\documentclass[12pt]{article}
\usepackage{epsf}

 \font\blackboard=msbm10 
 \font\blackboards=msbm7 \font\blackboardss=msbm5
 \newfam\black \textfont\black=\blackboard
 \scriptfont\black=\blackboards \scriptscriptfont\black=\blackboardss
 \def\Bbb#1{{\fam\black\relax#1}}

\def\d{{\rm d}}
\def\mi{{\rm i}}
\def\e{\mathop{\rm e}\nolimits}
\def\defi{\stackrel{\rm def}{=}}
\newcommand\grsim{\mathrel{\hbox{\lower1ex\hbox{\rlap{$\sim$}\raise1ex\hbox{$>$}}}}}
\newcommand\losim{\mathrel{\hbox{\lower1ex\hbox{\rlap{$\sim$}\raise1ex\hbox{$<$}}}}}

\title{``Exact WKB integration" of polynomial 1D Schr\"odinger
(or Sturm--Liouville) problem}
\author{{\bf Andr\'e Voros} \\
\\
CEA, Service de Physique Th\'eorique de Saclay\\
F-91191 Gif-sur-Yvette CEDEX (France)\\
{ E-mail : {\tt voros@spht.saclay.cea.fr}}\\
\\
and\\
\\
Institut de Math\'ematiques de Jussieu--Chevaleret\\
CNRS UMR 7586\\
Universit\'e Paris 7\\
2 place Jussieu,
F-75251 Paris CEDEX 05 (France)}

\begin{document}
\maketitle

We review an ``exact semiclassical" resolution method
for the general stationary 1D Schr\"odinger equation
\begin{equation}
\label{SCH}
-\hbar^2 \psi''(q) + [ V(q) + \lambda ] \psi (q)=0
\qquad (q \in \Bbb R)
\end{equation}
with a {\sl polynomial\/} potential
\begin{equation}
\label{POL}
V(q) \equiv +q^N+v_1 q^{N-1}+ \cdots + v_{N-1} q , \qquad
\mbox{denoting } (v_1, \cdots v_{N-1}) = \vec v.
\end{equation}
(In quantum mechanics, $\hbar\ (>0)$ is Planck's constant,
and the eigenvalue parameter $-\lambda \equiv E$ is the energy.)

{\sl Semiclassical\/} methods are small-$\hbar$ treatments or expansions
for quantum problems such as eq.(\ref{SCH}).
Generically, they deliver factorially divergent series in $\hbar$,
which at best qualify as asymptotic approximations
for the true quantum solutions.
Nevertheless, a pioneering work of Balian--Bloch \cite{BB} has established
that fully exact semiclassical formalisms are also achievable in principle.
This idea was then concretely implemented into an exact WKB method
for 1D problems like eq.(\ref{SCH}), using resurgence analysis
and (generalized) Borel resummation of $\hbar$-expansions \cite{V1}--\cite{VZ}; 
in this approach, however, a gap still remains in the proof 
of a required resurgence theorem (\cite{DP}, thm 1.2.1 and its Comment), 
and moreover, some of the most explicit calculation steps were reached 
only for very special cases \cite{V2}--\cite{V4}.

The subject of this review is a newer exact semiclassical (WKB) method
\cite{V0}--\cite{V7}
more directly based upon functional relations over the original parameter space,
in the line of Sibuya's framework \cite{S}.
This treatment is easier to formulate at once in broad generality ---
still within {\sl ordinary\/} (linear) differential equations
(i.e., 1D quantum problems).
Moreover, this approach reveals striking unexplained coincidences
(currently centered on {\sl homogeneous\/} differential equations)
with {\sl Bethe-Ansatz\/} methods
for some classes of exactly solvable {\sl 2D problems\/}:
certain statistical-mechanical lattice models
and conformal quantum field theories \cite{DT}--\cite{SU}.

As far as we know, exact semiclassical formalisms can be modeled on 
old-fashioned (asymptotic) ones.
Our own ``cooking recipe", in fact, is to ``exactify" (rewrite in exact form)
a conventional semiclassical framework: here, a {\sl WKB formalism\/}.
We will further specialize to a {\sl polarized\/} Sturm--Liouville problem,
given by eq.(\ref{SCH}) on a {\sl half-line\/} only:
$[0,+\infty)$ without loss of generality,
and with a choice of {\sl Neumann or Dirichlet\/} boundary conditions at $q=0$.
Accordingly, we will recall some basic WKB results (Sec.1),
then the exact ingredients that we incorporate (Sec.2),
and thereafter how they together combine into an exact semiclassical formalism,
first for eigenvalue problems (Sec.3),
and finally for eq.(\ref{SCH}) in full generality (Sec.4).
We intend to stress the governing ideas throughout, at the expense of
technicalities which received emphasis in our previous works.

\section{Standard asymptotic inputs}

Our essential starting semiclassical ingredient is
{\sl asymptotic WKB theory\/} in one dimension:
specifically, solutions of eq.(\ref{SCH})
(``wave functions", in quantum language) admit the WKB approximations
\begin{equation}
\label{wkb}
\psi_\pm (q) \sim \Pi_\lambda(q)^{-1/2}
\e^{ \pm {1 \over \hbar} \int^q \Pi_\lambda(q') \d q'} , \qquad \mbox{where}
\end{equation}
\begin{equation}
\label{PI}
\Pi_\lambda(q) \defi \sqrt{V(q) + \lambda}
\qquad \mbox{(= classical forbidden-region momentum)}.
\end{equation}
Moreover, as is well known, eq.(\ref{wkb}) holds
in a compound asymptotics sense:
not only for $\hbar \to 0$, but also, when the potential is polynomial,
at fixed $\hbar$ for $|\Pi_\lambda(q)| \to \infty$,
which includes $|\lambda| \to \infty$ {\sl or\/} $|q| \to \infty$, at will.
Henceforth, we choose to scale out $\hbar$ to unity.
Then, two distinct asymptotic consequences can be drawn from eq.(\ref{wkb}).

\subsection{Wave-function asymptotics ($q \to +\infty$)}

We first introduce a ``classical" $q \to +\infty$ expansion, with the notation
\begin{equation}
\label{BET}
[V(q) +  \lambda]^{-s+ 1/2} \sim
\sum_\sigma \beta_\sigma (s;\vec v,\lambda) q^{\,\sigma-Ns} \quad
\bigl( \sigma={N \over 2},\ {N \over 2}-1, {N \over 2}-2, \ldots \bigr) ;
\end{equation}
we will mostly need the leading $\beta_\sigma$ with $\sigma \ge -1$
(which are independent of $\lambda$, with one singular exception: when $N=2$~!)
and at $s=0$, which we then denote $\beta_\sigma(\vec v)$.

We now return to the quantum problem: following Sibuya \cite{S},
a {\sl recessive\/} or {\sl subdominant\/} solution $\psi_\lambda (q)$
of eq.(\ref{SCH}) on the half-line can be fully specified
by a decaying WKB behavior like (\ref{wkb}) in the $q \to +\infty$ direction.
The reexpansion of that WKB expression in $q$ using eq.(\ref{BET}) yields
\begin{equation}
\label{QAS}
\psi_\lambda (q) \sim {\mathcal C} q^{-N/4 \, -\beta_{-1}(\vec v)}
\exp \Bigl\{- \!\! \sum_{\{\sigma>-1\}} \!\! \beta_\sigma (\vec v)
{q^{\sigma+1} \over \sigma+1} \Bigr\} \qquad (q \to + \infty) ;
\end{equation}
only a constant normalization factor ${\mathcal C}$ remains to be assigned,
(e.g., ${\mathcal C} \equiv 1$ in \cite{S}).
Then, the solution $\psi_\lambda$ gets uniquely fixed
through the condition (\ref{QAS}).

\subsection{Eigenvalue asymptotics ($\lambda \to -\infty$)}

Here we consider the potential $V(|q|)$, thus extended to the whole real line
by mirror symmetry.
It is classically {\sl confining\/}, and gives rise to a
discrete quantum energy spectrum $\mathcal E$ (for the variable $-\lambda$).
Let the eigenvalues be $\{ E_k \}_{k=0,1,2,\ldots}$ in increasing order:
then, $E_k \uparrow +\infty$,
and the parity of the index $k$ moreover matches the parity 
of the eigenfunctions;
equivalently, even (resp. odd) $k$ label the Neumann spectrum ${\mathcal E}^+$
(resp. Dirichlet spectrum ${\mathcal E}^-$) on the half-line $[0,+\infty)$.

Now, the WKB approximation (\ref{wkb}) for the eigenfunctions entails
a semiclassical formula for $E=E_k$,
called {\sl Bohr--Sommerfeld quantization condition\/}:
\begin{equation}
\label{BSC}
\oint_{\{p^2 + V (|q|)=E \} } {p \,\d q \over 2 \pi} \sim k + {1 \over 2} \quad
\mbox{for integer } k \to + \infty ,
\end{equation}
valid asymptotically for large $k$, i.e., $E \to +\infty$.
If that condition is further expanded to all orders in $k^{-1}$,
then reorganized in descending powers of $E$, it takes the form \cite{V0}:
$E=E_k$ obeys
\begin{equation}
\label{BAS}
\sum_\alpha b_\alpha(\vec v) E^\alpha \sim k + {1 \over 2} \quad
\mbox{for integer } k \to + \infty \quad
\bigl( \alpha= \mu,\ \mu-{1 \over N},\ \mu-{2 \over N}, \cdots \bigr),
\end{equation}
where the leading exponent is $\mu \defi {1 \over 2} + {1 \over N}$
(with $ b_\mu E^\mu \equiv (2\pi)^{-1} \oint_{\{p^2+|q|^N=E\}} p \d q$,
the classical action for the homogeneous $V$ case).
Notes: the $b_\alpha(\vec v)$ are polynomial in the 
$\{ v_j \}_{j \le (\mu - \alpha)N}$,
and actually parity-dependent as well (unless $V$ is an even polynomial)
but only from $\alpha = -3/2$ downwards;
proper quantum corrections (beyond eq.(\ref{BSC})) only contribute 
to the orders $\alpha \le -\mu$, which are not critically needed here.

\section{Exact inputs}

We now explain how the preceding WKB approach can be made exact
by incorporating a single exact ingredient, essentially:
that the Wronskian of two solutions of eq.(\ref{SCH}) is a constant!
(precisely, it will be the computable, nonzero Wronskian 
of two particular solutions).

\subsection{Spectral determinants}

The main exact spectral function we will seek
is the canonical or {\sl zeta-regularized\/} determinant of the eigenvalues
$D(\lambda)$, formally $ \prod_k (\lambda + E_k)$.
It can be rigorously defined in terms of a spectral Hurwitz-zeta function,
$\sum_k (\lambda + E_k)^{-s}$, as an entire function of order $\mu$
in the variable $\lambda$ (and implicitly entire in $\vec v$), through \cite{VZ}
\begin{equation}
\label{DET}
\log D(\lambda) \defi
- \partial_s \biggl[ \sum_k (\lambda + E_k)^{-s} \biggr]_{s=0} .
\end{equation}

We will actually need the {\sl fixed-parity\/} determinants $D^\pm(\lambda)$
(for the separate spectra ${\mathcal E}^\pm$),
plus {\sl exact limit formulae\/} (of Euler--Maclaurin type)
which {\sl effectively\/} rebuild each determinant from its spectrum, as
\cite{V0}
\begin{eqnarray}
\label{DPM}
\log D^\pm(\lambda) \equiv \lim_{K \to + \infty} \Biggl \{
\sum_ {k<K} \log (E_{k}+\lambda) \!\!\!\! &+& \!\!\!\! 
{1 \over 2} \log (E_{K}+\lambda) \Biggr . \\
(\mbox{for } k,K \ {\textstyle{\rm even \atop odd}}) \qquad \qquad
&-& \!\!\!\!{1 \over 2} \sum_{ \{\alpha >0 \} } \Biggl. b_\alpha {E_K}^\alpha
\biggl[ \log E_K - {1 \over \alpha} \biggr] \Biggr \} . \nonumber
\end{eqnarray}
Here the second line expresses {\sl counterterms\/},
which have a semiclassical character:
each of them corresponds to a term of the {\sl diverging\/} (as $E \to \infty$)
initial part of the expansion (\ref{BAS}),
so as to achieve a finite $K \to +\infty$ limit.
Eq.(\ref{DPM}) amounts to limit-product representations of $D^\pm$,
which are {\sl rigidly determined by the zeros alone\/}
(unlike Hadamard products, which retain extra free factors).

\subsection{The basic exact identities}

By suitably integrating the Schr\"odinger equation along the half-line,
we can obtain a pair of identities relating the spectral determinants
to a subdominant solution $\psi_\lambda(q)$: 
\begin{equation}
\label{ID}
D^-(\lambda) \equiv \psi_\lambda(0), \qquad D^+(\lambda) \equiv
-\psi'_\lambda(0),
\end{equation}
subject to this explicit specification for 
the normalizing factor ${\mathcal C}$ in eq.(\ref{QAS}):
\begin{equation}
\label{NCT}
\log {\mathcal C} \equiv {1 \over N} \Bigl[ - (2 \log 2) \beta_{-1}(s ; \vec v)
+ \partial_s \Bigl( {\beta_{-1}(s ; \vec v) \over 1-2s} \Bigr) \Bigr]_{s=0} .
\end{equation}
Eqs.(\ref{ID},\ref{NCT}) answer a ``central connection problem",
as they provide some kind of data at $q=0$
for a solution that was fixed by its $q \to +\infty$ behavior.

Remarks: in \cite{V0,V6}, we erroneously presumed that ${\mathcal C}$ 
always equals unity,
whereas this is only guaranteed when
$\beta_{-1}(s,\vec v) \equiv 0$ (which happens not so rarely, however);
the general case was fixed in \cite{V0} (Corrigendum) and \cite{V7}.
In the latter reference, the result (\ref{NCT}) was tied to the fact that
``classical determinants" $D_{\rm cl}^\pm(\lambda)$ can also be defined,
again by eq.(\ref{ID}) but replacing the exact $\psi_\lambda$ 
by its WKB approximation (from eq.(\ref{wkb})).

\subsection{The conjugate problems}

As with roots of an algebraic equation (which are better studied all together),
we will not treat the Schr\"odinger equation (\ref{SCH}) in isolation,
but together with a properly defined set of ``conjugate" equations,
involving the continuation of eq.(\ref{SCH})
to the whole {\sl complex domain\/},
and analytic dilations controlled by an angle $\varphi \defi 4\pi/(N+2)$.

The {\sl first conjugate\/} Schr\"odinger equation is eq.(\ref{SCH})
with potential $V^{[1]}$ and spectral variable $\lambda^{[1]}$,
defined as \cite{S}
\begin{equation}
\label{CNJ}
V^{[1]}(q) \defi \e^{-\mi\varphi} V(\e^{-\mi\varphi/2}q),
\qquad \lambda^{[1]} \defi \e^{-\mi\varphi} \lambda .
\end{equation}
The iterated rotations (by $\ell\varphi$) likewise generate a sequence of
conjugate equations, which repeats itself with a {\sl finite period\/} $L$ 
given by
\begin{equation}
\label{NCJ}
L \equiv \left\{ { N+ 2 \quad \mbox{generically} \hfill} \atop
\displaystyle {N \over 2}+ 1 \quad \mbox{for {\sl even\/} polynomials } V(q) .
\right.
\end{equation}
The distinct conjugate equations are thus labeled by 
$[\ell]=\ell\ {\rm mod}\,L$, and likewise for all related quantities: 
e.g., the $(N-1)$-uplet of coefficients $\vec v$ from eq.(\ref{POL})
induces a sequence of conjugates $\vec v^{[\ell]}$.

\subsection{The Wronskian identity}

Now comes the key exact input, derivable from all preceding considerations.

Let $\psi_\lambda^{[1]}(q)$ be the recessive (for $q \to +\infty$) solution
identically defined for the first conjugate equation; 
then, by counter-rotation (analytic dilation),
$ \Psi_\lambda (q) \defi \psi_{\lambda^{[1]}}^{[1]} (\e^{\mi\varphi/2}q) $ is 
a solution to the original equation (\ref{SCH}) decaying in an 
{\sl adjacent\/} Stokes sector relative to the first solution $\psi_\lambda(q)$.
Here, {\sl no Stokes phenomenon has to be resolved\/}:
both solutions can be asymptotically tracked in a common direction,
e.g., $q \to +\infty$, 
and their (constant) Wronskian extracted in this limit \cite{S},
giving 
\begin{equation}
\label{WRP}
\Psi_\lambda (q)\psi'_\lambda (q)-\Psi'_\lambda (q)\psi_\lambda (q)
\equiv 2{\mathcal C}^{[1]} {\mathcal C}\,\mi
\e^{\mi\varphi/4} \e^{\mi\varphi\beta_{-1}(\vec v)/2} .
\end{equation}
If the basic identities (\ref{ID}) are substituted into (\ref{WRP}),
with ${\mathcal C}^{[1]} \equiv 1/{\mathcal C}$ following from eq.(\ref{NCT}),
this Wronskian formula also reads as
\begin{equation}
\label{DW}
\e^{+\mi\varphi/4}
D^+( \e^{-\mi\varphi} \lambda ; \vec v^{[1]} ) D^-( \lambda ; \vec v)
-\e^{-\mi\varphi/4}
D^+( \lambda ; \vec v) D^-( \e^{-\mi\varphi} \lambda ; \vec v^{[1]} )
\equiv 2 \mi \e^{\mi\varphi\beta_{-1}(\vec v)/2}
\end{equation}
(reinstating the parametric dependence of the determinants upon $\vec v$).

In standard approaches, this explicit Wronskian is viewed
as an elementary but auxiliary input,
and the main dynamical information is sought in {\sl Stokes multipliers\/},
namely Wronskians of {\sl nonadjacent\/} solution pairs.
For instance, a typical ``basic" Stokes multiplier 
(one involving a next-to-adjacent solution pair)
has a formula (unnormalized) analogous to eq.(\ref{DW}),
\begin{equation}
\label{SM}
C(\lambda ; \vec v) \equiv {1 \over 2\mi}
\bigl[ \e^{+\mi\varphi/2} D^+( \lambda ; \vec v)
D^-( \e^{+2\mi\varphi} \lambda ; {\vec v}^{[-2]})
-\e^{-\mi\varphi/2} D^+( \e^{+2\mi\varphi} \lambda ; {\vec v}^{[-2]})
D^-( \lambda ; \vec v) \bigr] .
\end{equation}
However, such multipliers are accessible only through connection formulae
involving {\sl Stokes phenomena\/}, 
hence in general they are not explicitly known.

By contrast, our claim is that complete exact information can be squeezed
out of the {\sl elementary Wronskian\/} (\ref{WRP}) or (\ref{DW}) alone.

\section{Exact analysis of eigenvalue problem}

We now argue the  sufficiency of eq.(\ref{DW}) as exact input to recover
the {\sl spectral determinants\/} (or the eigenvalues of $V(|q|)$).
Remark: such an exact WKB analysis for the {\sl Airy functions\/}, 
using the potential $|q|$ \cite{V4}, was a {\sl basic milestone\/} 
to understand the general case:
its validity testified that {\sl a kink singularity is harmless\/} 
in the potential (although it looks pathological 
from a naive complex-WKB viewpoint,
i.e., worse than any complex-analytic singularity).

At first sight, 
the single functional equation (\ref{DW}) seems definitely incomplete
to determine the two unknown functions $D^\pm( \lambda ; \vec v)$.
Formally, its general solution should allow one function to remain arbitrary.
Concretely, we may even write the full system of all
$L$ distinct conjugates of eq.(\ref{DW}), 
then solve it either for $D^+$ in terms of $D^-$, or vice-versa, 
but in any case one function out of the pair will stay wholly unconstrained.

Remark: invoking Stokes multipliers is of no help in this perspective;
for instance, eq.(\ref{SM}) adds one equation indeed,
but also one unknown function, $C( \lambda ; \vec v)$ itself.

\subsection{An exact quantization condition}

First of all, we shift to the {\sl spectra\/} as basic unknowns,
given that the determinants are fully recoverable from them
through eq.(\ref{DPM}).
For instance, let us seek the odd spectrum ${\mathcal E}_-$
(the even spectrum can be separately processed likewise).
Setting $\lambda=-E_k$ with $k$ odd already reduces eq.(\ref{DW})
to a multiplicative form:
\begin{equation}
\label{MLT}
-\e^{-\mi\varphi/4}
D^+( -E_k ; \vec v) D^-( -\e^{-\mi\varphi} E_k ; \vec v^{[1]} ) =
2 \mi \e^{\mi\varphi\beta_{-1}(\vec v)/2} \qquad (k=1,3,5,\ldots).
\end{equation}
Then, dividing this by its first conjugate formula,
we manage to eliminate the even spectrum altogether, obtaining
\begin{figure}
{\hfill
\epsfysize=9cm
\epsfbox{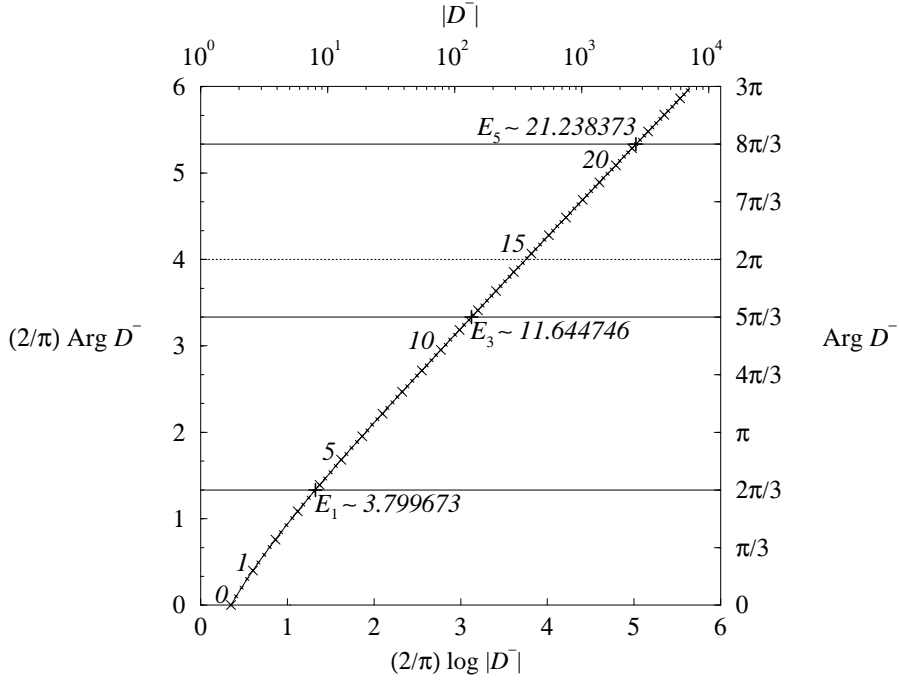}
\hfill}
\caption{\small
The exact quantization formula (\ref{EQM}) for the odd eigenvalues 
of the homogeneous quartic potential, $V(q)=q^4$: the curve
$ \{ D^-(\e^{\mi\pi/3} E ; \vec 0) \}_{ \{E > 0\} }$
intersects the straight line 
$ \{ {\rm Arg}\, D^- (\lambda ; \vec 0) = 2 \pi/3 \ {\rm mod}\, \pi \} $
precisely at $E=E_k$, with $k$ odd 
(complex-logarithmic coordinates are used; 
the $E$-scale along the curve is labeled in italics).}
\end{figure}
\begin{equation}
\label{MUL}
{D^-(-\e^{-\mi\varphi} E_k ; {\vec v}^{[+1]} ) \over
D^-(-\e^{+\mi\varphi} E_k ; {\vec v}^{[-1]} )}
= -\e^{+\mi\varphi/2\,+\, \mi\varphi\beta_{-1}(\vec v)}
\qquad (k=1,3,5,\ldots).
\end{equation}
Finally, by taking logarithms
(fixed by analytical continuation from the origin in $(E,\vec v)$-space),
we end up with a formula fulfilled by the eigenvalues,
which has an outer {\sl Bohr--Sommerfeld form\/} (namely, $F(E)=k+1/2$)
but is {\sl completely exact\/}: for $k$ odd, $E=E_k$ obeys 
\begin{equation}
\label{EQM}
{2 \over \pi} {\rm Arg\,} D^-(-\e^{-\mi\varphi}E;{\vec v}^{[+1]})
-{\varphi \over \pi} \beta_{-1}(\vec v)
= k+{1 \over 2} - {N-2 \over 2(N+2)}
\qquad  (k=1,3,5,\ldots).
\end{equation}

This is an exact spectral formula for ${\mathcal E}_-$ indeed...
but in terms of (the determinant corresponding to) another similar spectrum,
the first conjugate ${\mathcal E}_-^{[1]}$.
So, this step still ends with two unknowns (spectra)
for a single constraint (the quantization condition (\ref{EQM})).

Special example (homogeneous case): if $V(q)=|q|^N$, 
then $\beta_{-1}(0; \vec 0, \lambda) \equiv \delta_{N,2} \,\lambda/2$,
all conjugate spectra coincide, 
and the left-hand side of eq.(\ref{EQM}) reduces to 
$(2/\pi) {\rm Arg\,} D^-(-\e^{-\mi\varphi}E;{\vec 0}) + \delta_{N,2}\, E/2$;
fig.1 illustrates this in the case $N=4$.

\subsection{Complete set of exact quantization conditions}

Next, we observe that each of the conjugate potentials $V^{[\ell]}(q)$
for given $\ell$ can be handled wholly in parallel to the original potential.
Consequently, its spectra ${\mathcal E}_\pm^{[\ell]}$ also obey
exact quantization conditions extending eq.(\ref{EQM}) and its even-$k$ analog.
They are now complex in general, and involve
the {\sl two adjacent spectra\/} (of the same parity)
${\mathcal E}_\pm^{[\ell+1]}$ and ${\mathcal E}_\pm^{[\ell-1]}$, as
\begin{eqnarray}
\label{EQC}
-{\mi \over \pi} \Bigl[ \log D^\pm(-\e^{-\mi\varphi}E;{\vec v}^{[\ell+1]})
-\log D^\pm(-\e^{+\mi\varphi}E;{\vec v}^{[\ell-1]}) \Bigr]
- (-1)^\ell \,{\varphi \over \pi} \beta_{-1}(\vec v) =
\nonumber \\
=  k+{1 \over 2} \pm {N-2 \over 2(N+2)} \qquad \mbox{for } E=E_k^{[\ell]},\quad
k={\textstyle{0,2,4,\ldots \atop 1,3,5,\ldots}}
\end{eqnarray}
to which {\sl must\/} be appended all the conjugates of eq.(\ref{DPM}),
\begin{eqnarray}
\label{DC}
\log D^\pm(\lambda ;{\vec v}^{[\ell]}) = \lim_{K \to +\infty} \Biggl \{
\sum_ {k<K} \log (E_k^{[\ell]} + \lambda)
\!\!\!\! &+& \!\!\!\! {1 \over 2} \log (E_K^{[\ell]} + \lambda) \Biggr . \\
(\mbox{for } k,K \ {\textstyle{\rm even \atop odd}}) \qquad \qquad
&-& \!\!\!\! {1 \over 2} \sum_{ \{\alpha >0 \} } \Biggl.
b_\alpha({\vec v}^{[\ell]}) \bigl[ E_K^{[\ell]} \bigr] ^\alpha
\biggl[ \log E_K^{[\ell]} - {1 \over \alpha} \biggr] \Biggr \} . \nonumber
\end{eqnarray}
The latter serve to eliminate the determinants from eqs.(\ref{EQC});
moreover, eqs.(\ref{DC}) imply {\sl extra conditions\/} 
through the coefficients $b_\alpha({\vec v}^{[\ell]})$,
namely the {\sl asymptotic\/} (Bohr--Sommerfeld) formulae
extending eq.(\ref{BAS}) to all conjugate spectra,
\begin{equation}
\label{BAC}
\sum_\alpha b_\alpha({\vec v}^{[\ell]}) \bigl[ E_k^{[\ell]} \bigr] ^\alpha
\sim  k + {1 \over 2} \quad \mbox{for integer } k \to + \infty :
\end{equation}
these {\sl have to be obeyed\/}, simply for the definiteness of eqs.(\ref{DC}).
But the present exact quantization conditions (\ref{EQC}) are far too implicit
to include the semiclassical ones, eqs.(\ref{BAC}), as a limit:
like the Wronskian identity (\ref{DW}) (of which they are just an offspring),
eqs.(\ref{EQC}) explicitly incorporate only the degree $N$ 
plus one invariant, $\beta_{-1}(\vec v)$, out of the whole dynamics
(they are ``quasi-universal");
whereas the totality of the coefficients $v_j$ of the potential are reflected 
in eqs.(\ref{BAC}) (we expand these at least over $\{ \alpha > -\mu \}$),
and nowhere else.
So, the asymptotic Bohr--Sommerfeld formulae (\ref{BAC}) 
need to be {\sl asserted independently\/} here;
they now come as crucial {\sl boundary conditions\/}
(for $k=+\infty$) to eqs.(\ref{EQC}).

Remark: for the difference term in brackets (left-hand side of eq.(\ref{EQC})),
eqs.(\ref{DC}) reduce to a {\sl convergent series\/} (except for $N=1$ or 2), 
namely $ \sum_k \bigl[ \log (E_k^{[\ell]} - \e^{-\mi\varphi}E)
- \log (E_k^{[\ell]} - \e^{+\mi\varphi}E) \bigr] $.
The counterterms are then optional
(but still very helpful to accelerate the convergence of this series).

Now, because the conjugates form a finite cyclic set,
each full system (\ref{EQC}$^\pm$) (for a fixed parity)
precisely gives $L$ exact equations for $L$ unknown spectra.
Equivalently, we have reached a potentially complete and closed system
for each global unknown 
$ {\mathcal E}_+^\bullet $ or $ {\mathcal E}_-^\bullet $,
which we define as:
$ {\mathcal E}_\pm^\bullet \defi \bigcup_{[\ell]}
\bigl( \{ \ell \} \times \e^{\mi\ell\varphi} {\mathcal E}_\pm^{[\ell]} \bigr) $,
the labeled union of all conjugate spectra
(rotated by $\e^{\mi\ell\varphi}$ for sheer convenience).

Just as the homogeneous-potential cases correspond to 2D integrable models
and their Bethe-Ansatz exact ``solutions" \cite{DT}--\cite{SU},
the exact quantization conditions (\ref{EQC}--\ref{BAC}) can also be viewed as
{\sl Bethe-Ansatz equations for the general polynomial-potential case\/}.
In the same line,
we may then say that the Schr\"odinger eigenvalue problem (\ref{SCH})
gets ``integrated" by that Bethe-Ansatz system,
presuming that the latter can be solved uniquely and effectively.

The challenge posed by the eigenvalue problem in eq.(\ref{SCH}) hence gets
displaced onto the new system (\ref{EQC}--\ref{BAC}),
which is of a very different nature:
it is {\sl nonlinear\/}, {\sl selfconsistent\/} (or bootstrapping),
and above all, purely discrete, {\sl no longer differential at all\/}
--- albeit still (countably) infinite-dimensional.
Such elaborate novel features make us unable to further discuss 
the solvability of that system on rigorous terms
(even for the simplest cases of homogeneous potentials, e.g., $V(q)=q^4$);
e.g., we cannot answer basic questions such as:

- can one compute backwards: do the exact quantization conditions (\ref{EQC})
imply the Wronskian identities ((\ref{DW}) and its conjugates)
for the determinants reconstructed by eq.(\ref{DC}),
and perhaps even the eigenvalue property (\ref{SCH})~?

- can the system (\ref{EQC}--\ref{BAC}) have a (reasonably?) unique solution;
optimally, can it be turned into a fixed-point problem
for some {\sl contractive mapping(s)\/}, in the form
${\mathcal M}^\pm \bigl\{ {\mathcal E}_\pm^\bullet \bigr\} =
{\mathcal E}_\pm^\bullet$~?

\subsection{Numerical tests}

We are nevertheless able to tackle the resolution of eqs.(\ref{EQC}--\ref{BAC})
{\sl numerically\/}.
This system can be approximated on computer
by $K$-dimensional mapping schemes (with $K$ large).
Our implementation exploits a ``division of labor" which appears to exist
within that system (though not as sharply as we now state it):
for $k \to +\infty$ ($k \grsim K$ numerically), 
where asymptotic semiclassical theory is accurate enough to be used alone,
the eigenvalues $E_{k}^{[\ell]}$ get essentially determined by 
the explicit {\sl asymptotic\/} Bohr--Sommerfeld formulae (\ref{BAC}),
and this enacts the boundary conditions;
the finite-$k$ (or $k \losim K$) eigenvalues 
then adjust themselves selfconsistently in response to 
the implicit {\sl exact\/} quantization conditions (\ref{EQC}), 
to which {\sl they\/} are more sensitive.

We have thus tested homogeneous potentials (up to $q^{400}$) 
on the one hand \cite{V2}--\cite{V4}, 
and moderately inhomogeneous quartic and sextic potentials
on the other hand \cite{V0}.
The results strongly support the most favorable picture proposed above, 
namely that nonlinear dynamics of eqs.(\ref{EQC}--\ref{BAC}) can be contractive.
Under straightforward iterations, 
our numerical mappings indeed displayed geometric convergence,
and towards the correct quantum spectra (up to finite-$K$ effects).
Within each fixed degree $N$, the homogeneous case looked the most contractive.
For quartic potentials $V(q)=q^4+vq^2$, 
we have validated such convergence in the range $-10 \le v \le +5$:
as we decreased $v$ away from 0 (double-well regime), we quit at $v=-10$
after seeing nothing worse than
a graceful numerical degradation (a gradual decrease of final accuracy);
by contrast, in the opposite direction (single-well regime),
iteration schemes could easily be made stable for moderate $v$,
but each of them (so far) rather abruptly ``derailed" into numerical instability
at some point, $v \approx +5$ being our highest result.
So, we cannot yet see the expected smooth transition
towards the harmonic regime as $v \to +\infty$ within those calculations.

\subsection{Analytical generalizations}

The numerical difficulty just mentioned led us to further study
the case $V(q)=q^4+vq^2$ purely analytically.
The $v \to \infty$ regime is singular indeed as it induces a discontinuity
(a jump from 4 to 2) in the degree $N$,
the controlling factor of the exact analysis.
But far from revealing any intrinsic singularities,
the exact WKB formalism proved regular
also in the large-$v$ regime (\cite{V7}, Sec.3).
For instance, it supplied the
large-$v$ behavior for the (zero-energy) spectral determinants
${\rm Qi}^\pm (v) \defi {\det}^\pm  \bigl( -\d^2 /\d q^2 + q^4 + vq ^2 \bigr)$:
\begin{equation}
\label{QD2}
{\rm Qi}^\pm (v) \sim
\e^{-v^{3/2}/3} {\det}^\pm  \bigl( -\d^2 /\d q^2 + vq ^2 \bigr)
\quad \Bigl[ \equiv \e^{-v^{3/2}/3} v^{\pm 1/8}
{\sqrt {2 ^{1 \pm 1} \pi} \over
\Gamma \bigl( {\textstyle{2 \mp 1 \over 4}} \bigr)} \Bigr],
\end{equation}
plus many exact results for these functions ${\rm Qi}^\pm (v)$
(mostly around $v=0$,
whereas standard perturbation theory operates around $v = +\infty$).

Extension is straightforward to general binomial potentials, i.e., 
to the functions $\det^\pm \bigl(-\d^2 /\d q^2 + q^N + vq ^M \bigr)\quad (M<N)$.
The $v$-roots of those determinants solve 
the {\sl generalized eigenvalue problems\/}
\begin{equation}
\label{GEP}
- \psi''(q) + [ q^N + vq ^M ] \psi (q)=0 ,
\end{equation}
in which the coupling $v$ is now taken as spectral parameter
whilst the original one is frozen ($\lambda \equiv 0$),
and eq.(\ref{GEP}) is posed over $\{q>0\}$ just as before,
with Neumann or Dirichlet conditions at $q=0$.
The cases where $N=2M+2$ are special; for even $M$ and selected $v$, 
these give {\sl supersymmetric\/} potentials.
Exact WKB analysis still works here (\cite{V7}, Sec.4; \cite{SU}):
it yields exact quantization conditions for $v$,
which degenerate to a totally explicit form when $N=2M+2$ 
(compare with eq.(\ref{EQC}) for $\ell=0$):
\begin{equation}
\label{EQNM}
\Bigl[ -{\varphi \over \pi} \beta_{-1}(\vec v) \equiv \Bigr]
\ -{2 \over N+2} v = k+{1 \over 2} \pm {N-2 \over 2(N+2)}
\qquad \mbox{for } k= \textstyle{0,2,4,\ldots \atop 1,3,5,\ldots} .
\end{equation}
The usual harmonic oscillator spectral problem,
formerly an isolated exception, is now seen as the first in
an infinite sequence of solvable cases ($N=2,6,10,\ldots$).

A {\sl wholly novel\/} exactly solvable family is also uncovered
by that exact WKB analysis,
namely the generalized spectrum of eq.(\ref{GEP})
{\sl over the whole real line\/} when $N=2M+2$ with $M$ {\sl odd\/}.
(For even $M$, this spectrum just consists of both parity components
of eq.(\ref{EQNM}).)
For odd $M$, the full spectral determinant over the whole real line comes out as
\cite{V7}
\begin{equation}
\label{DTR}
\det \bigl( -\d^2 /\d q^2 + q^{2M+2} + vq ^M \bigr) \equiv
{\cos \pi \nu v \over \sin \pi \nu} , \qquad
\nu \defi {1 \over N+2} \equiv  {1 \over 2M+4}
\end{equation}
implying the exact quantization condition
\begin{equation}
\label{EQNE}
v = (2M+4)(n+1/2) \quad (n \in \Bbb Z) \qquad \mbox{if $M$ is odd.}
\end{equation}
Here, even the lowest case $M=1$ does not reduce to a classic solvable case.
The corresponding eigenfunctions have also been recently investigated \cite{BW}.

Furthermore, both degenerate quantization conditions (\ref{EQNM},\ref{EQNE})
get identically reproduced
by the leading semiclassical Bohr--Sommerfeld formula (\ref{BSC}):
i.e., as for the harmonic oscillator, ``semiclassical quantization is exact"
(but for those generalized eigenvalues only).

\section{Exact wave-function analysis}

\subsection{A solution algorithm for the Schr\"odinger equation}

We finally show how the exact treatment can solve the differential equation
(\ref{SCH}) itself,
i.e., compute the recessive solution $\psi_\lambda(q)$ for instance,
in just a slight extension of the previous exact eigenvalue calculations:
it suffices to use the basic identities (\ref{ID}) {\sl in reverse\/}.
The choice of origin being immaterial, those can be written over the half-line
$[q,+\infty)$ as
\begin{equation}
\label{DI}
\psi_\lambda (q) \equiv D_q^-(\lambda + V(q)), \qquad
\psi'_\lambda (q) \equiv -D_q^+(\lambda + V(q))
\end{equation}
where $D_q^\pm(\lambda)$ are the spectral determinants of the potential
$V_q(\cdot)$ defined by $[V(\cdot)-V(q)]$ over the half-line $[q,+\infty)$,
and by mirror symmetry on the complementary half-line.

Thus, for every fixed $q$, $\psi_\lambda (q)$ can be computed from
the Dirichlet spectrum ${\mathcal E}_{q,-}$ of the potential $V_q(\cdot)$
(and likewise, $\psi'_\lambda$ from the Neumann spectrum).
That spectrum in turn obeys an exact quantization formula, now $q$-dependent,
hence can derive from an {\sl explicit fixed-point equation\/}
of the form ${\mathcal M}_q^- \bigl\{ {\mathcal E}_{q,-}^\bullet \bigr\} =
{\mathcal E}_{q,-}^\bullet$.
Once this spectrum has been attained, possibly by iteration,
one extra application of the reconstruction formula like eq.(\ref{DPM})
for the determinant $D_q^-$ can supply any value of this function,
in particular the one required by eq.(\ref{DI}).

We emphasize that all distinct points $q$ get handled completely rigidly
and autonomously here, invoking no propagation from one point to the next.
In this sense, the whole procedure that we have described ``integrates"
(one solution for) the Schr\"odinger equation (\ref{SCH}).

\subsection{Numerical tests}

\begin{figure}
\vskip-.3cm
\epsfysize 8cm
{\hfill
\epsfbox{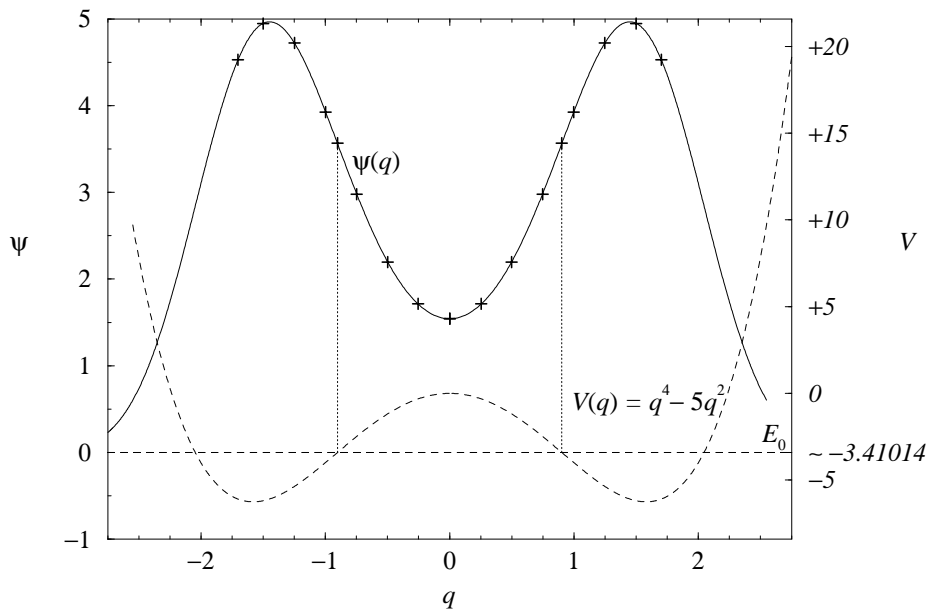}
\hfill}
\caption{\small 
Exact-WKB calculations for a ground-state eigenfunction $\psi(q)$, 
in the potential $V(q)=q^4-5q^2$.
{\bf Dashed curves} (using the right-hand vertical scale): 
the potential $V(q)$ and its ground-state energy level $E_0 \approx -3.41014$,
to show the classically allowed ($\{ V(q) \le E_0 \}$)
and forbidden ($\{V(q)>E_0 \}$) portions of the $q$-axis;
vertical lines mark the two inner turning points.
{\bf Continuous curve} (using the left-hand vertical scale): 
eigenfunction $\psi(q)$ resulting from a computer integration 
of eq.(\ref{SCH}) (using the NAG routine D02KEF),
and normalized according to eqs.(\ref{QAS},\ref{NCT}) 
with ${\mathcal C} = 1$ here (which is not $L^2$ normalization!).
{\large \tt +}~: exact-WKB values $\psi(q)$,
reached by iterative calculations of the determinants $D_q^-$ of eq.(\ref{DI}),
{\sl independently for each sampled $q$-value\/}
(numerical instability disrupts this calculation scheme beyond
$|q| \approx 1.7$).}
\end{figure}

We have validated this exact WKB algorithm numerically upon several examples
with quartic potentials, and for solutions $\psi(q)$ in fully quantum regimes;
we provide two illustrations.

Fig.2 displays a {\sl ground-state wave function\/}
(for a symmetric double-well potential, featuring $\beta_{-1} \equiv 0$).
Now, inasmuch as the present approach works for all recessive ($q \to +\infty$)
solutions alike, it also cannot directly detect which ones will end up being
square-integrable for $q \to -\infty$ as well;
the easiest way to pin down an eigenfunction is then  
to preassign $-\lambda = E$ itself as the eigenvalue.
So, the ground-state eigenvalue $E_0\ (\approx -3.41014)$ 
is determined separately first,
e.g., by means of the basic exact quantization conditions (\ref{EQC});
the resulting number $\lambda = -E_0$ is then fed into the calculation
of every value $\psi_\lambda (q)$ according to eqs.(\ref{DI})
and the associated $q$-dependent quantization conditions.
As in the eigenvalue problem, a numerical instability abruptly sets in 
outside a parameter range which is $\{ |q| \losim 1.7 \}$ here
(but the contracting factors of our iterations are numerically
$\approx 0.67$ for most of the computed $q$-points,
and do not exceed 0.86 at the farthest computable ones).
Furthermore, the stability range encompasses two {\sl turning points\/}
($q \approx \pm 0.90267$), which clearly do not affect the exact WKB calculations.

\begin{figure}
{\hfill
\epsfysize=8cm
\epsfbox{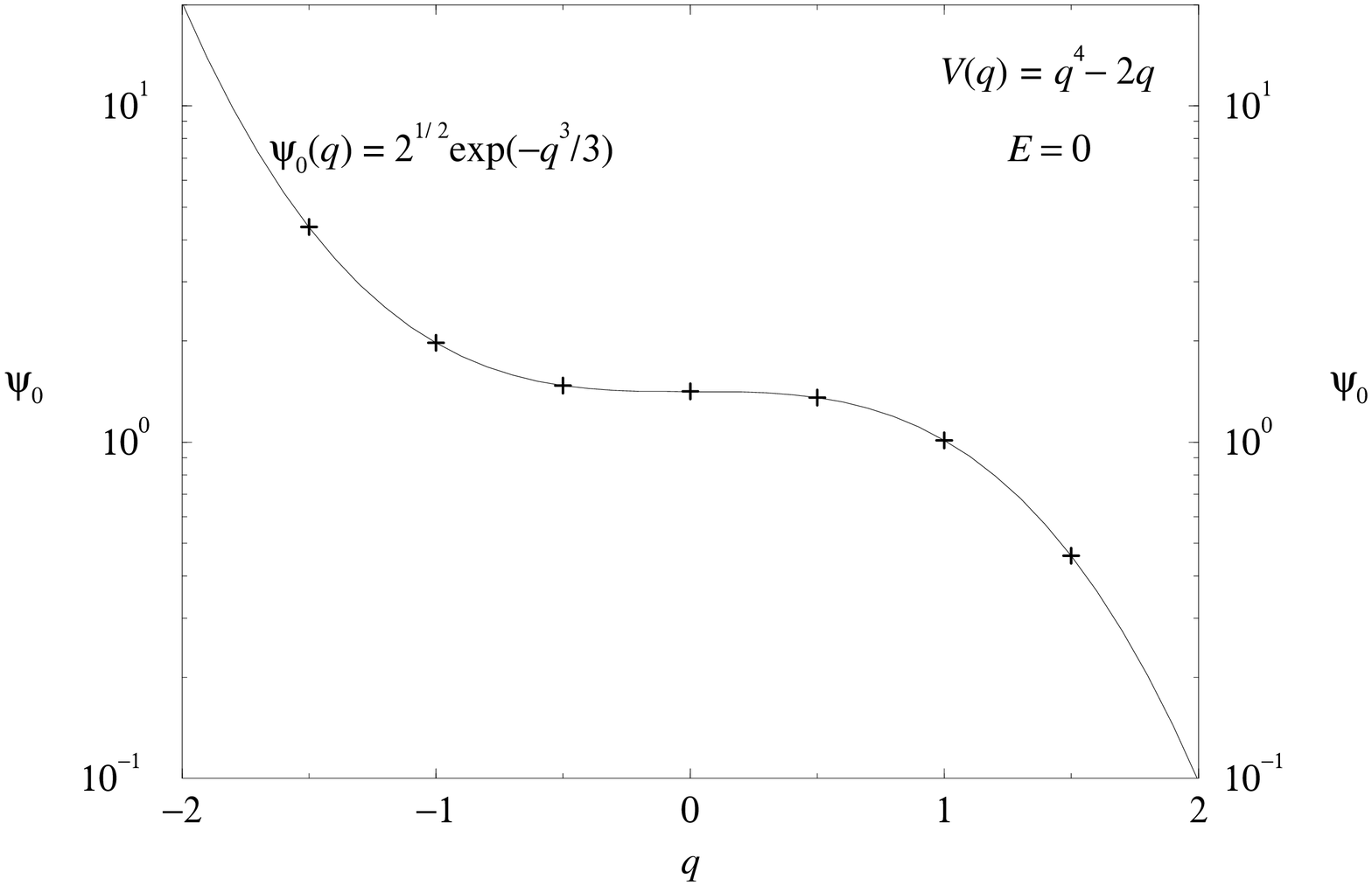}
\hfill}
\caption{\small
Exact-WKB calculations for a non-$L^2$ recessive solution
$\psi_0(q)$, in the potential $V(q)=q^4-2q$ and at energy $E=0$.
{\bf Continuous curve} (now using a logarithmic vertical scale): 
analytical solution $\psi_0(q)$, normalized according to
eqs.(\ref{QAS},\ref{NCT}) with ${\mathcal C} = \sqrt 2$ here.
{\large \tt +}~: exact-WKB computed values $\psi_0(q)$, as in fig.2.}
\end{figure}

Fig.3 shows a contrasting example: a {\sl non-square-integrable\/} solution, 
now for an asymmetrical quartic potential 
featuring a nonzero $\beta_{-1}(s ; \vec v)$ $(\equiv 2s-1$).
The chosen problem is eq.(\ref{GEP}) with $N=4,\ M=1$, and with $v=-2$
{\sl not\/} belonging to the generalized spectrum (given by eq.(\ref{EQNE})):
in this problem, a supersymmetry (on $\{q>0\}$ with Neumann boundary conditions)
makes the fully normalized solution available in closed form.
The iterations again display contraction factors $\approx 0.67$ 
(at all plotted points).

\subsection{Concluding remarks}

We first restate our two major open issues (which might be fruitfully merged).
On the one hand, the nonlinear infinite-dimensional structure of 
eqs.(\ref{EQC}--\ref{BAC}) is entirely uncontrolled at a rigorous level.
On the other hand, the analytical coincidences observed 
with exactly solvable 2D theories \cite{DT}--\cite{SU} are in utter want
of fundamental explanations (and, hopefully, of generalizations too).

We are also confident that such approaches will extend 
beyond polynomial potentials.
On the basis of existing results (\cite{DT}, resp. \cite{ZD}),
we believe that exact-WKB resolution methods should next be sought for 
rational, resp. trigonometric-polynomial, potentials;
a later realistic and desirable target could be all Heun-class equations.

\end{document}